\documentclass[a4paper,10pt]{article}

\newcommand{\n}{\nu}
\newcommand{\m}{\mu}
\newcommand{\s}{\sigma}
\def\equ#1{(\ref{#1})}
\title{Antisymmetric tensor matter fields in a curved space-time}
\author{C.A.S. Silva and R.R. Landim}

\begin{document}

\maketitle

\begin{abstract}
An analysis about the antisymmetric tensor matter fields
Avdeev-Chizhov theory in a curved space-time is performed. We
show, in a curved spacetime , that the Avdeev-Chizhov theory can
be seen as a kind of a $\lambda\varphi^{4}$ theory for a  "complex
self-dual" field. This relationship between Avdeev-Chizhov theory
and $\lambda\varphi^{4}$ theory  simplify the study of tensor
matter fields in a curved space-time. The energy-momentum tensor
for  matter fields is computed.
\end{abstract}

\section{Introduction}
Antisymmetric tensor fields have been introduced since many years
and are object of continuous and renewed interests due to their
connection with the topological field theories \cite{d.birmingham-prt209, gt.horowitz-cmp125, gt.horowitz-cmp130, m.blau-anp205}.
In 1994 L.V. Avdeev and M.V. Chizhov \cite{lv.avdeev-plb321}
achieved the construction of a four dimensional abelian gauge
model which includes antisymmetric second rank tensor fields as
matter fields rather than gauge fields. The model contains also a
coupling of the antisymmetric fields with chiral spinors and a
quartic tensor self-interaction term. It exhibits several features
among which we underline the asymptotically free ultraviolet
behavior of the abelian gauge interaction. More recently V.
Lemes, R.R. Landim and S.P. Sorella \cite{v.lemes-plb352} found the
interesting result that the tensor matter invariant lagrangian
proposed for Avdeev-Chizhov can be seen as a kind of a
$\lambda\varphi^{4}$ theory for a complex antisymmetric field
satisfying a "complex self-dual condition" in Minkowski spacetime,
give us a straightforward way of obtaining its nonabelian
generalization.\\
In this work we shall to generalize the Avdeev-Chizhov theory for
curved space-times. It would be a very tiresome task. Therefore we
shall begin this analysis studying the relationship between
Avdeev-Chizhov theory and $\lambda\varphi^{4}$ theory in a curved
space-time. We shall show that the equivalence between
Avdeev-Chizhov theory and $\lambda\varphi^{4}$ theory archived for
\cite{v.lemes-plb352} in a flat space-time remains if a curvature
is introduced. As we shall see, It will simplify the study of
tensor matter fields in a curved space-time.
\section{Complex self-dual condition in a curved space-time}
The self-dual complex condition introduced in
\cite{v.lemes-plb352} depend on signature and dimension of
space-time. In this section we shall investigate the self-dual
complex condition in a four dimensional curved space-time with
signature
-2, which is used in general relativity. \\
The dual of a tensor $F_{\m\n}$ is defined as:
\begin{equation}
\tilde{F}_{\mu\nu}=\frac{1}{2}\epsilon_{\mu\nu\rho\sigma}F^{\rho\sigma},\label{dualF}
\end{equation} \\
with
\begin{eqnarray}
\epsilon_{\mu\nu\rho\sigma}=\sqrt{-g}\varepsilon_{\mu\nu\rho\sigma}
\;\;\;\;\;\; ; \;\;\;\;\;
\epsilon^{\mu\nu\rho\sigma}=\frac{1}{\sqrt{-g}}\varepsilon^{\mu\nu\rho\sigma},\label{levicivitacurvo}
\end{eqnarray} 
where $\varepsilon^{\mu\nu\rho\sigma}$ is the Levi-Civita
tensor.\\

Let us build the complex field $\varphi_{\m\n}$ like in the flat
case \cite{v.lemes-plb352}:\\
\begin{equation}
\varphi_{\mu\nu} = T_{\mu\nu}+ i\widetilde{T}_{\mu\nu}.
\end{equation}\\
Then, we have:
\begin{eqnarray*}
\widetilde{\varphi}_{\mu\nu} = \widetilde{T}_{\mu\nu}+\emph{i}
\widetilde{\widetilde{T}}_{\mu\nu},
\end{eqnarray*}
where $\widetilde{T}_{\m\n}$ is given for \equ{dualF} and
\begin{equation}
\widetilde{\widetilde{T}}_{\m\n} = -T_{\m\n}.
\end{equation}
Hence, we have
\begin{eqnarray*}
\widetilde{\varphi}_{\m\n} = \widetilde{T}_{\m\n} - iT_{\m\n} \rightarrow
i \widetilde{\varphi}_{\m\n} = T_{\mu\nu}+\emph{i}\tilde{T}_{\mu\nu} = \varphi_{\m\n}
\end{eqnarray*}
or
\begin{equation}
i\widetilde{\varphi}_{\mu\nu}= \varphi_{\mu\nu}.
\label{auto-dual-complex}
\end{equation}\\
Then we can build complex self-dual fields in a four dimensional
curved space-times with signature -2.
\section{Tensor matter fields as a $\lambda\varphi^{4}$ theory in a curved space-time}
We shall show now that the Avdeev-Chizhov theory for matter fields
can be seen as a kind of $\lambda\varphi^{4}$ theory in a curved
space-time for a self-dual complex field $\varphi_{\m\n}$.
Firstly, let us write the action for complex field
$\varphi_{\m\n}$ in Minkowski space-time:
\begin{eqnarray}
S_{inv} &=&\int\!d^{4}x -\frac{1}{4g^{2}}F_{\m\n} F^{\m\n}+ i \bar{\psi} \gamma^{\m}
\partial_{\m}\psi - \bar{\psi} \gamma^{\m}\gamma_{5}A_{\m}\psi \\ \nonumber \vspace{4mm}
&-&\int\!d^{4}x \Big( \, (\varphi^{\m\n}_{\;\;| \m })^{*}(\varphi_{\sigma\n}^{\;\;| \sigma})
+ \frac{q}{8} (\varphi^{*\m\n} \varphi_{\n\alpha}\varphi^{*\alpha\beta}\varphi_{\beta\m}) \, \Big) \\ \nonumber
&+&\int\!d^{4}x\frac{1}{2}y \bar{\psi} \sigma_{\m\n}(\varphi^{*\m\n} + \varphi^{\m\n})\psi,            \label{acao-completa}
\end{eqnarray}
\noindent where $\varphi_{|}$ indicates the gauge covariante derivative of $\varphi$ field \cite{m.carmeli-cfgrgt},which we go to define as
\begin{equation}
\varphi_{\m\n \;|\; \alpha} = -i(\varphi_{\m\n \; , \; \alpha} - 2iA_{\alpha}\varphi_{\m\n}),
\end{equation}
\noindent what is justified by the fact $\varphi_{|}$ transforms under gauge transformations \equ{gauge-transforms} as the fields $\varphi$. 

The action \equ{acao-completa} is invariant under the
infinitesimal abelian gauge transformations:
\begin{eqnarray}
&\delta&\!\!\!\!\! A_{\m} = \partial_{\m}\omega  \;\;\;\;\; , \;\;\;\;\;
\delta \psi = -i\omega \gamma_{5} \psi \ , \\ \nonumber \vspace{8mm}
&\delta&\!\!\!\! \bar{\psi} = -i\omega \bar{\psi} \gamma_{5} \;\;\;\;\; , \;\;\;\;\;\;
\delta \varphi_{\m\n} = -2\omega \varphi_{\m\n} \ . \label{transformacao3}
\end{eqnarray}
Let us now, using the general covariance principle, perform the
following transformations:
\begin{eqnarray*}
\eta_{\m\n} &\rightarrow& g_{\m\n} \;\;\;\; ; \;\;\;\;
d^{4}x \rightarrow \sqrt{-g}d^{4}x\\ \\
\varphi^{\m\n}_{\;\ ,\ \alpha} &\rightarrow& \varphi^{\m\n}_{\;\;
;\ \alpha}\;\;\;\; ; \;\;\;\;
\varphi^{\m\n}_{\;\ |\alpha} \rightarrow \varphi^{\m\n}_{\;\;
||\alpha} \\ \\
\gamma^{\m} &\rightarrow& \rho^{\m}\;\; \;\; ; \;\;\;\;
\psi_{,\ \m} \rightarrow\psi_{;\ \m},
\end{eqnarray*}\\
\noindent where $\varphi_{||}$ indicates the gauge and coordinate covariante derivative of $\varphi$ field \cite{m.carmeli-cfgrgt}, which is defined as
\begin{equation}
\varphi_{\m\n \;||\; \alpha} = -i(\varphi_{\m\n \; ; \; \alpha} - 2iA_{\alpha}\varphi_{\m\n}) 
\end{equation}
 
Let us also include the gravitational action\\
\begin{eqnarray*}
S_{g} = \int\!\!\!\sqrt{-g}Rd^{4}x
\end{eqnarray*}\\
where $R = g_{\m\n}R^{\m\n}$ is the Ricci scalar. \vspace{4mm} \\
Then, the action for the $\varphi_{\m\n}$ field in a curved
space-time is
:\\
\begin{eqnarray}
S_{inv} = \int\!\!\!\sqrt{-g}d^{4}x(\!\!\!&\frac{1}{16\pi G}R
-\frac{1}{4g^2}F_{\m\n} F^{\m\n}+ i \rho^{\m} \psi_{;\ \m} -
\bar{\psi} \rho^{\m}\gamma_{5}A_{\m}\psi
-(\varphi_{\|\m}^{\m\n})^{*}
{(\varphi^{\|\sigma}_{\sigma\n})} \nonumber\\
&+ \frac{q}{8} ( {\varphi^{*\m\n}} \varphi_{\n\alpha}
{\varphi^{*\alpha\beta}}
         \varphi_{\beta\m}) +
 \frac{1}{2}y \bar{\psi}\sigma_{\m\n}(\varphi^{*\m\n}+\varphi^{\m\n})\psi \  ). \label{acao-fi-curv}
\end{eqnarray} \\
The action \equ{acao-fi-curv} is, like in the flat case, invariant under the infinitesimal abelian gauge transformations: :
\begin{eqnarray}
&\delta&\!\!\!\!\!\!\!A_{\m} = \omega_{;\ \m} = \omega_{,\ \m}
\;\;\;\;\; ; \;\;\;\;\; \delta \psi = -i\omega \gamma_{5} \psi \\
\nonumber \vspace{8mm} &\delta&\!\!\!\!\!\!\bar{\psi} = -i\omega
\bar{\psi} \gamma_{5} \;\;\;\;\;\;\;\;\;\;\; ; \;\;\;\;\; \delta
\varphi_{\m\n} = -2\omega \varphi_{\m\n}. \label{gauge-transforms}
\end{eqnarray}  \vspace{-3mm}

\noindent Writing $\varphi_{\m\n}$ as
\begin{equation}
\varphi_{\m\n} = T_{\m\n} + i\widetilde{T}_{\m\n},
\end{equation}
and using the following identities : \vspace{-7mm}

\begin{eqnarray*}
\widetilde{T}_{\m\lambda} \widetilde{T}^{\lambda\n} = T_{\m\lambda}
T^{\lambda\n} + \frac{1}{2} \delta^{\n}_{\m} T_{\alpha\beta}
T^{\alpha\beta},
\end{eqnarray*}
\begin{eqnarray}
 T_{\m\lambda}\widetilde{T}^{\lambda\n} = -\frac{1}{4}
\delta^{\n}_{\m} {\ } T_{\alpha\beta} \widetilde{T}^{\beta\alpha},
\end{eqnarray}
\begin{eqnarray}
 \widetilde{T}_{\m\rho}\widetilde{T}^{\sigma\rho}_{;\ \sigma} = -\frac{1}{2}
          T^{\alpha\beta} T_{\alpha\beta;\ \m}
     - (T_{\lambda\m;\ \sigma}) T^{\sigma\lambda},
\end{eqnarray}
\begin{eqnarray}
\widetilde{T}_{\alpha\beta} T^{\alpha\beta;\ \m}_{\;\;\;\;\;\;\;\;\ ;\
\m} =
  -2 T_{\alpha\beta}\widetilde{T}^{\n\alpha;\ \beta}_{\;\;\;\;\;\;\;\;\ ;\ \n}
  -2 \widetilde{T}_{\alpha\beta}T^{\n\alpha;\ \beta}_{\;\;\;\;\;\;\;\;\ ;\ \n},
\end{eqnarray} \vspace{1cm}
\noindent after a straightforward calculation, we thus arrive at
the following action
\begin{eqnarray*}
S_{inv} = \int \!\!\!\sqrt{-g}d^{4}x(\!\!\!&\frac{1}{16\pi G}R
-\frac{1}{4g^2} F_{\m\n} F^{\m\n} + i \rho^{\m} \psi_{;\ \m} -
\rho^{\m}\gamma_{5}A_{\m}\psi + \frac{1}{2} {(T_{\m\n;\
\lambda})}^2 - 2 {(T_{\m\n}^{\;\;\ ;\
\m})}^2 \\ \\
&+ 4 A_{\m}(T^{\m\n}\widetilde{T}_{\lambda\n}^{\;\;\ ;\ \lambda} -
\widetilde{T}^{\m\n}T_{\lambda\n}^{\;\;\ ;\ \lambda}) + 4 (\frac{1}{2}
{(A_{\lambda}T_{\m\n})}^2 -2 {(A^{\m}T_{\m\n})}^2) \\ \\
&+y \bar{\psi} \s_{\m\n} T^{\m\n} \psi + \frac{q}{4}(
\frac{1}{2}{(T_{\m\n}T^{\m\n})}^2 -
               2 T_{\m\n}T^{\n\rho}T_{\rho\lambda}T^{\lambda\m})\ )  ,
\end{eqnarray*}

\noindent and
\begin{eqnarray}
&\delta&\!\!\!\!\!\!A_{\m} = \omega_{;\ \m} = \omega_{,\ \m} \;\;\;\; ; \;\;\;\;
\delta \psi = -i\omega \gamma_{5} \psi \\ \vspace{8mm} \nonumber
&\delta&\!\!\!\!\!\!\bar{\psi} = -i\omega \bar{\psi} \gamma_{5} \;\;\;\; ; \;\;\;\;
\delta T_{\m\n} = -2\omega \widetilde{T}_{\m\n}.
\end{eqnarray}
The action above is a Avdeev-Chizhov like action in a curved
space-time.
\section{Energy-momentum tensor}
In this section we shall to compute the energy-momentum tensor for antisymmetric matter fields. As we shall see, the relationship between Avdeev-Chizhov and $\lambda\varphi^{4}$ theory simplify this.\\
The energy-momentum tensor is very important in general relativity
for describe the matter distribution which acts as sources for the
gravitational field. It is defined as \cite{m.carmeli-cfgrgt}
\begin{equation}
\Theta_{\m\n} =
\frac{2}{\sqrt{-g}}\Big\{{\frac{\partial(\sqrt{-g}L)}{\partial
g^{\m\n}} - \frac{\partial}{\partial
x^{\alpha}}[\frac{\partial(\sqrt{-g}L)}{\partial g^{\m\n}_{,\
\alpha}}]}\Big\}. \label{tensor momentum-energia}
\end{equation}
The direct use of Avdeev-Chizhov Lagrangian in \equ{tensor
momentum-energia} would lead to very tiresome computations. Then
let us use the Lagrangian for $\varphi_{\m\n}$ rather the
Lagrangian for $T_{\m\n}$ field.\\
Firstly, let us analyze the second term of right side of \equ{tensor
momentum-energia}, which is what it gives more work to calculate. We have that only the kinetic term of
\equ{acao-fi-curv} could depend on derivatives of the metric in relation to the coordinates.
However, using the self-dual complex condition
\equ{auto-dual-complex} and the following identity:
\begin{eqnarray}
\widetilde{\varphi}^{\alpha\beta}_{\;\ ||\ \alpha} =
\frac{1}{2}\epsilon^{\alpha\beta\m\n}\varphi_{\m\n|\ \alpha},
\end{eqnarray}\\
we have that:\\
\begin{eqnarray*}
(\varphi^{\m\n}_{\;\ ||\ \m})^{*}(\varphi_{\s\n}^{\;\ ||\ \s})& = & g^{\s\omega}g^{\rho\phi}g^{\tau\gamma}\epsilon^{\m\n\alpha\beta}\epsilon_{\omega\n\phi\gamma}(\varphi_{\alpha\beta\ |\m})^{*}(\varphi_{\rho\tau\ |\s})\\
&=&\frac{-1}{4}g^{\s\omega}g^{\rho\phi}g^{\tau\gamma}\delta^{\m\alpha\beta}_{\omega\phi\gamma}(\varphi_{\alpha\beta\ |\m})^{*}(\varphi_{\rho\tau\ |\s}).\\
\end{eqnarray*}\\
As we see, the Lagrangian for $\varphi^{\m\n}$ field don't depend on metric derivatives.
Hence the second term of right side of \equ{tensor momentum-energia} vanish.\\
Computing the other term we found for energy-momentum tensor:
\begin{eqnarray}
\Theta_{\m\n} &=& 3g^{\rho\phi}(\varphi_{[\phi\n \; , \omega]})^{*}(\varphi_{\rho\m}^{\;\; , \;\omega} - \frac{1}{2}g^{\tau\omega}\varphi_{\rho\tau\; ,\; \m})\\  \nonumber &-&\frac{q}{8}(\varphi^{*\lambda\gamma}\varphi_{\gamma\m}\varphi_{\n}^{* \beta}\varphi_{\beta\lambda} + 3\varphi^{*\lambda\gamma}\varphi_{\gamma}^{\; \sigma}\varphi^{*}_{\sigma\n}\varphi_{\m\lambda})\\  &-& \frac{1}{2}g_{\m\n}L + (\m\leftrightarrow\n). \label{tensor momentum energia ii}
\end{eqnarray}\\
In terms of matter field $T_{\m\n}$ we have:
\begin{eqnarray*}
\Theta_{\m\n} = &-& (3/2)g^{\rho\phi}T_{[\omega\phi \;\; , \;
\n]}(T_{\rho\m}^{\;\; , \; \omega} -
\frac{1}{2}g^{\tau\omega}T_{\rho\tau\;\; , \;\m}) + ( T \rightarrow \widetilde{T} )\\ &+&
\frac{q}{2}(T_{\m}^{\; \lambda}T_{\n\lambda}T^{\omega\phi}T_{\omega\phi} + 4T_{\beta\n}T_{\m\lambda}T^{\lambda\alpha}T_{\alpha}^{\; \beta})\\ &-& \frac{1}{2}g_{\m\n}L +
(\m\leftrightarrow\n).
\end{eqnarray*}\\

If we include the interactions of tensor matter fields with eletromagnetic field, we shall to find
\begin{eqnarray}
\Theta_{\m\n} &=& 3g^{\rho\phi}(\varphi_{[\phi\n \ |\; \omega]})^{*}(\varphi_{\rho\m}^{\;\ |\;\omega} - \frac{1}{2}g^{\tau\omega}\varphi_{\rho\tau\ |\; \m})\\  \nonumber &-&\frac{q}{8}(\varphi^{*\lambda\gamma}\varphi_{\gamma\m}\varphi_{\n}^{* \beta}\varphi_{\beta\lambda} + 3\varphi^{*\lambda\gamma}\varphi_{\gamma}^{\; \sigma}\varphi^{*}_{\sigma\n}\varphi_{\m\lambda})\\ &+& \frac{1}{4h^{2}}\sqrt{-g} \Big( \frac{1}{4}g_{\m\n}F_{\alpha\beta}F^{\alpha\beta} - F_{\m\alpha}F_{\n}^{\; \alpha} \Big) - \frac{1}{2}g_{\m\n}L + (\m\leftrightarrow\n). \label{tensor momentum energia ii.i}
\end{eqnarray}
It's easy to verify that $\Theta_{\m\n}$ is invariant under gauge transformations \equ{gauge-transforms}
\begin{equation}
\delta\Theta_{\m\n} = 0.
\end{equation}

\section{Conclusion}
We have shown here that Avdeev-Chizhov theory for antisymmetric
tensor matter fields can been seen as a kind of
$\lambda\varphi^{4}$ theory for a self-dual complex field in a
curved space-time. Also we have shown that the relationship above
simplifies the computation of energy-momentum tensor, fundamental
object in general relativity, for matter fields in a curved
space-time, making the study of Avdeev-Chizhov theory in a curved
space-time easier.
\section{Acknowledgements}
We wish to thank the Conselho Nacional de Desenvolvimento
Cient\'ifico e Tecnol\'ogico, CNPQ-Brazil and Funda\c{c}\~ao Cearense de
Apoio ao Desenvolvimento Cient\'ifico e Tecnol�gico-Cear\'a, Brazil
for the financial support.

\end{document}